# Theory and computation of thermal-field emission from semiconductors

Salvador Barranco Cárceles[1*], Veronika Zadin[2],
Aquila Mavalankar[3], Ian Underwood[1], Andreas Kyritsakis[2]

[1]School of Engineering, University of Edinburgh, Scotland
[2]Institute of Technology, University of Tartu, Estonia
[3]Adaptix Imaging Ltd., Oxford, England
[*]Corresponding author: salvador.barranco-carceles@univ-lyon1.fr, s.barranco.carceles@gmail.com

ABSTRACT

Semiconducting field emitters present some interesting features (e.g.; self-limited electron emission) for both scientific interest and industrial applications. The analysis of experimental results and device design has been restrained by the lack of accurate 3D models for the simulation of thermal-field emission from semiconductors. Here we review and correct the equations of field emission from semiconductors and include them to expand GETELEC (General Tool for Electron Emission Calculations). Our method covers all electron emission regime (field, thermal, and intermediate), aiming to maximise the calculation accuracy while minimising the computational cost. GETELEC-2.0 is able to reproduce the characteristic non-linear I-V curves in Fowler-Nordheim coordinates obtained from semiconductors, giving insights about their nature. As well as providing an explanation to the lack of experimental observation of valence band electrons from semiconductors.

*Key works:* electron emission, semiconductors, simulations

## I. Introduction

Semiconducting field emitters (s-FE) combine unique and appealing features such as self-limited emission [1] and superior photo-electron yield [2]. These make s-FE desirable for high specification technologies like information displays [3], ultrashort pulsed photon sources [4], and portable 3D medical imaging devices [5].

The widespread commercialisation of field emission-based products remains elusive. This is due to the knowledge gap between field emission theory and experimental observations [6] and the resulting limited understanding about the emitter's surface dynamics, which makes the development of field emitter-based devices a challenging task. This gap is more acute for s-FE given their higher complexity (presence of field penetration and surface states) when compared to metallic field emitter (m-FE).

Extensive work has been done to develop a model for field emission [7] and vacuum break down [8] in m-FE. Both models have provided a quantitative understanding of the behaviour of metallic surfaces under high electrical fields ($> 5$ GV/m). However, the fundamental study of field emission from semiconductors date back to 1980 [9], [10], [11], [12]. Those studies are limited to 1D flat surfaces, do not consider the state-of-the-art field emission theory, nor the Nottingham and Joule heat effects on the emission. Therefore, the data interpretation from recent semiconducting electron emitters (e.g., sharp emitters with tip radius < 20 nm [13] or nanowires [14]) is expected to be inaccurate.

Accurate modelling and analysis tools will enable researchers to better understand the behaviour of semiconducting emitters; thus, along with high-quality experimental data, bridging the knowledge gap between theory and experimental observations, and providing the basis for effective design and realisation of semiconducting field emission devices.

In this work we develop a numerical model for the evaluation of the electron emission current density, Nottingham heat, and electron spectra from semiconducting emitters, based on Stratton's formulation [9], [10], [15] of the emission integrals from the conduction and valence bands. Our model is based on the existing computational software GETELEC (1.0) [7], which we expand here and update it into GETELEC 2.0 to include thermal-field emission from semiconductors, while maintaining the incorporation of emitter curvature effects. Link to repository here.

GETELEC 2.0 allows the general and efficient calculation of the emission characteristics (current density, Nottingham heat, electron spectra), which we then study for a wide range of emission parameters, contextualising them to provide insights in the physics of semiconducting field emission devices.

The paper is organised as follows. In Section II, we give the theoretical framework of our model, deriving the mathematical expressions it uses (full derivation in the Appendix) and defining the relevant physical quantities and parameters. In Section III, we give a brief overview of the new numerical approach. In Section IV, we present calculation results that provide insights about the emission physics from semiconductors and discuss the advantages and limits of our model. We finalise the paper summarising our findings and stating future research lines.

## II. Theory and Methods

### A. General framework of the Physics of Field Emission from Semiconductors

Figure 1 shows the potential energy diagram of a n-type semiconductor under high electric field. The bottom of the conduction band ($E_C$), the top of the valence band ($E_V$), the band gap ($E_g$), the Fermi level ($E_F$), and the work function ($\phi = \chi + \zeta$) are given in eV. We set the vacuum energy ($E_{vac}$) at 0 eV, with negative energies going downwards into the band. $j_C$ and $j_V$ are the emitted current density from the conduction and valence band, respectively, in $A/nm^2$. The diagram also shows the potential barrier and its two components: the image potential (doted curved line) and the potential due to the external electric field (solid straight line).

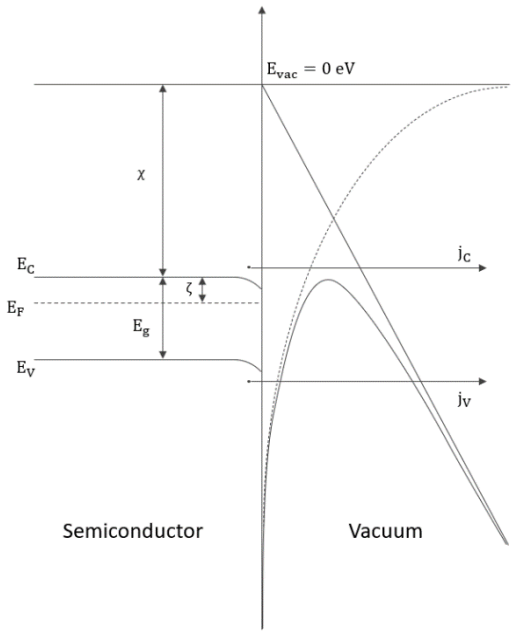

Figure 1: Energy diagram of a n-type semiconductor under high electric field for the zero current approximation ($E_F$ constant in all the semiconductor).

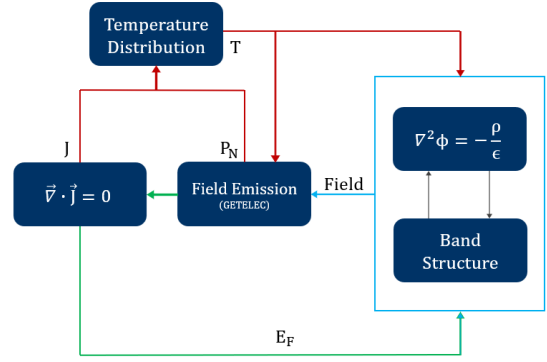

Figure 2: Schematic of the effects involved in semiconductor field emitters and their dependencies.

In metals, to a very good approximation and due to the large number of free charges, there is no potential drop in the emitter, i.e., the electric field does not penetrate the material. In semiconductors, in contrast, when a forward external field is applied (the emitter being the cathode), the field penetrates deep into the material bending the bands downwards as the electrons accumulate on the surface of the material, as shown in Figure 1. Therefore, the calculation of the electric field at the semiconductor surface requires a self-consistent solution of the Poisson's equation (Fig. 2 – blue loop) with the carrier concentration equations.

Upon the emission of a significant emission current density, the Fermi level also bends as the zero-current approximation becomes invalid [11]. To capture this change in $E_f$, another self-consistent loop (Fig. 2 – green loop) to solve the continuity equation needs to be added to the calculation. Finally, as the emission grows exponentially, the emitter's temperature rises due to Joule and Nottingham effects, which results in a broadening of the Fermi-Dirac distribution. This effect modifies the number of charges available and the potential drop at the surface (see the above argument); thus, changing the energy distribution of the emitted electrons. The nature of the emission transits from field driven (cold) to temperature driven (hot) as the temperature rises. Those effects require of a third self-consistent loop (Fig. 2 - red loop) that solves the head diffusion equation to accurately calculate the emission regime.

It is clear from the above analysis, that field emission from semiconductors poses a highly coupled and non-linear problem that requires a holistic self-consistent solution for the entire emitter geometry to yield accurate solutions. This results in computationally heavy calculations due to the high dimensionality of the problem and the lack of analytical solutions. The first step to tackle this challenge is developing a numerical model that allows for the evaluation of the emitted current and Nottingham heat for any point in the parameter space. In this paper we focus exactly on this first step. An analysis of the full self-consistent model of Figure 2 will be given on a separate forthcoming publication.

### B. Equations of emission from semiconductors

Stratton [9] provided a first set of equations for electron emission from the conduction and valence bands. We start our analysis from Eq. (4) [10], changing the nomenclature for clarity. Within the quasi-planar emitter approximation, we can treat the potential barrier as one-dimensional, thus ignoring the tangential dependence of the potential. Within this approximation, only the energy component normal to the emitting surface participates in the tunnelling. We define this energy component (normal components are denoted with the subscript "n" and the tangential components with the subscript "t") as:

$$E_n = E - E_t \qquad (1)$$

where E is the total energy of the electron and $E_t$ is the tangential energy, with components $E_x$ and $E_y$ (see Appendix Eq. A.19). Now we can calculate the emitted electron density from the conduction band by integrating over the momentum space in the band:

$$j_C = \frac{q}{4\pi^3 \hbar^3} \int f_{FD}(E) D(E - E_t) \frac{\partial E}{\partial p_n} d^3p \qquad (2)$$

where q is the electron charge, $\hbar$ is the reduced Plank constant, $f_{FD}(E)$ is the Fermi-Dirac distribution and $D(E)$ is the transmission coefficient given by the Kemble formula within the JWKB approximation [16]:

$$D(E_n) = \frac{1}{1 + e^{G(E_n)}} \qquad (3)$$

with

$$G(E_n) = g \int_{z_1}^{z_2} \sqrt{U(z) - E_n} \, dx \qquad (4)$$

and

$$U(z) = E_{vac} - qV(z) - \frac{Q}{z(1 + z/2R)} \qquad (5)$$

as on [7] (see equations 4 to 6). E and $E_t$ correspond to:

$$E = \frac{p^2}{2m^*} \qquad (6)$$

$$E_t = \frac{p_x^2 + p_y^2}{2m} = \frac{p_t^2}{2m} \qquad (7)$$

with $m^*$ as the effective and m as the free electron mass. Note that Eq. (7) originates from demanding the conservation of the transverse component of crystal momentum of the Bloch wave through the tunnelling process. In our view, whether the crystal momentum p or the actual momentum $m \partial_p E$ should be conserved is an open question that goes beyond the scope of this paper. Here we shall follow the standard approach in the literature of conserving p.

Conducting the transformations Eq. (1) results in (see Appendix for full development):

$$j_C = L \int_0^E l_{FD}(E_n)(D(E_n) - \bar{\alpha} D(\bar{\alpha} E_n)) dE_n \qquad (8)$$

where

$$L = \frac{qm}{2\pi^2 \hbar^3} \qquad (9)$$

$$\bar{\alpha} = 1 - \frac{m^*}{m} \qquad (10)$$

$$l_{FD}(E_n) = -\ln\left(\frac{1}{1 + e^{\frac{E_n - E_f}{k_b t}}}\right) \qquad (11)$$

$k_b$ is Boltzmann's constant (eV/K) and t in the temperature (K), with integration going from $E_c$ to $E_{vac}$. Eq. (8) is equivalent to Stratton's Eq. (7) [10] but with slightly different notation.

From an intermediate step in the development of Eq. (8), we can extract the electron energy distribution of the emitted electrons as:

$$J_C(E) = f_{FD}(E) \int_{\overline{\alpha}E}^{E} D(E_n)dE_n \quad (12)$$

The conduction band component of the Nottingham heat can be calculated as:

$$P_{N_C}(E) = \int_0^E J_C(E)(E - E_R)dE \quad (13)$$

where $E_R$ is the energy of the replacement electrons.

Similarly, as we did for the conduction band in Eq. (2) we derive the electron emission from the valence band as

Similarly, to the conduction band expression, we derive the current density for the valence band by performing integral manipulation in Eq. (14). It results in Eq. 15, which differs from Stratton's Eq. (A1) [15]. This difference arises from a mathematical mistake in Stratton's derivation (see Appendix for details).

$$j_V = \frac{q}{4\pi^3 \hbar^3} \int f_{FD}(\overline{E}) D_V(\overline{E} + \overline{E}_t) \frac{\partial \overline{E}}{\partial p_n} d^3p \quad (14)$$

[15][10]

$$j_V = L \int_{-\infty}^{E_V} l_{FD}(E_n)(D(E_n) - \overline{\alpha}D(\overline{\alpha}E_n - \alpha E_V))dE_n \quad (15)$$

where

$$\alpha = \frac{m^*}{m} \quad (16)$$

$$\overline{\alpha} = 1 + \frac{m^*}{m} \quad (17)$$

Similar as we did in Eq. (8) we can obtain the electron energy distribution for the valence band:

$$J_V(E) = f_{FD}(E) \int_{\overline{\alpha}E - \alpha E_V}^{E} D(E_n)dE_n \quad (18)$$

from which we can develop the valence band emission contribution to the Nottingham heat:

$$P_{N_V}(E) = \int_0^E J_V(E)(E_R - E)dE \quad (19)$$

### III. Code Implementation

The existing version of GETELEC (1.0) was based on using the appropriate formulae that can approximate both the values of the transmission coefficient $D(E_n)$ and the integrals of the previous sections for various regimes. The tunnelling barrier expression (Eq. (2.7) in [17]) can be approximated by the following empirical formula [7]:

$$U(z) = F \frac{R(\gamma - 1)z + z^2}{\gamma z + R(\gamma - 1)} \quad (20)$$

where F is the local electric field at the surface, R is the curvature parameter, and γ is a parameter that corresponds to the ratio of the far field to the local one. In a recent work [18], Kyritsakis showed that the barrier at any curved surface point can be approximated by an asymptotic expansion that is compatible with Eqs. (21-22), with R being the radius of *average* curvature at a given point. Equation (23) was specifically chosen to maintain this asymptotic behaviour for $z \ll R$, but also behave well at large distances (the parabolic form would turn upwards).

The calculation of the transmission coefficient was done by numerical integration of the JWKB formula based either on Eqs. (21-22), with the asymptotic formulae of [17] being used when applicable for efficiency.

Then, the emission was based on the GTF model [7], [19], i.e., distinguishing various regimes based on the field and temperature, and applying appropriate approximations. Using full numerical integration only in when it is absolutely necessary. The issue is that a full numerical integration of both the JWKB integral and over energies is too computationally costly to be used in conjunction with large-scale models where the emission needs to be evaluated many times over many emission points. On the other hand, these integrals become far more complex for semiconductors and GTF-like approximations are not available.

To overcome this roadblock, in GETELEC 2.0 we follow a different computational approach. Within the new approach, the potential barrier is conceptually treated as independent of the emitter material. The shape of the barrier depends on the three parameters F, R, and γ. The Gamow function

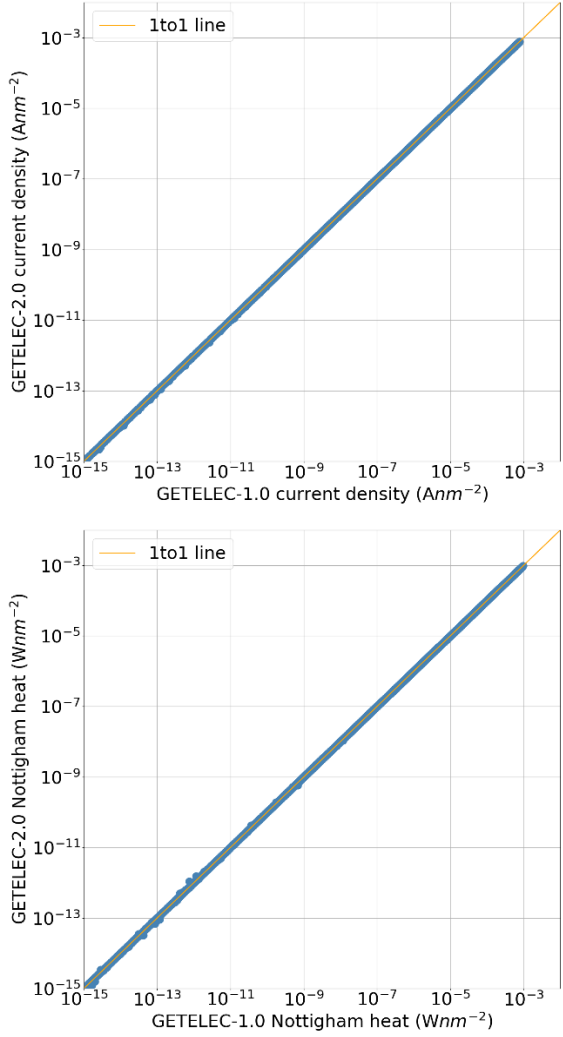

Figure 3: a) Current density calculation comparison between GETELEC-1.0 and GETELEC-2.0, with RMSE = 0.01531. b) Nottingham heat calculation comparison, with RMSE = 0.02987

$G(E_n)$ shall also depend on these parameters. Our computational approach is to pre-calculate the function $G(E_n; F, R, \gamma)$ by numerically integrating Eq. (4) for various values of $F, R,$ and $\gamma$ and fit the $G(E_n)$ function to a 4th order polynomial. Then, the coefficients of the polynomial along with the limits of validity of the polynomial approximation are tabulated and stored as an array. When a calculation is requested, the entire $D(E_n)$ function can be retried by interpolating the four polynomial coefficients, with its evaluation becoming very computationally efficient as it reduces to the evaluation of a 4th order polynomial and an exponential. This new method allows the evaluation of any of the emission equations of the previous section at a very affordable computational cost, offering the generality and accuracy of the full numerical integration.

To validate the model, we calculate j and $P_N$ by performing the full barrier calculation (GETELEC-1.0) and by the new computational approach (GETELEC-2.0). We calculated 8000 random combinations for F, R, T, ϕ, and γ over a broad range of values to ensure numerical accuracy and stability. Figure 3 shows the linear relation between the two methods for the range of current that is realistic to be measured on an experimental setup (J > 10 fA/nm$^2$).

Given a root-mean-square errors (RMSE) equal to 0.01531 and 0.02987 for the calculation of j and $P_N$ respectively, we are confident that our interpolation approach can be used as a general, accurate, and computationally efficient method to evaluate the emission characteristics, applicable to all regimes of thermal-field emission, and being easily extendible to photo-emission, as well as emission from materials with complex supply functions (e.g., semi-metals, molecules on surfaces, superconductors, etc.).

## IV. Results

### 1. Electron energy distribution

GETELEC-2.0 has been used to simulate the field emission properties of intrinsic Germanium (i-Ge) with $E_C$ = -4.4 eV, $E_F$ = -4.75 eV, $E_g$ = 0.7 eV, $m^*_e$ = 0.98m kg and $m^*_h$ = 0.59m kg, where m is the free electron mass. F = 5 GV/m, R = 20 nm, and γ = 10.

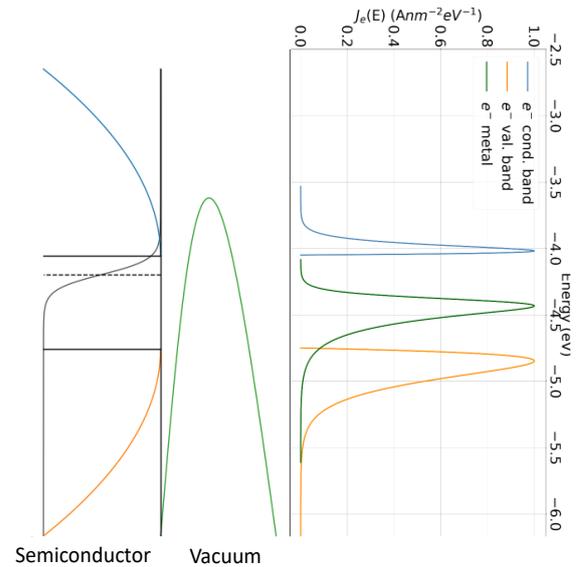

Figure 4: a) Density of States in blue for the conduction band and orange for the valence band. Fermi-Dirac distribution in black. Potential barrier in green. b) normalised electron energy distribution, with electrons from conduction band in blue, from valence band in orange and from a metal in green.

Three band positioning have been simulated: (1) flat band as shown in Figure 4a, (2) electron accumulation ($E_C = E_F$), and (3) semiconductor degeneracy ($E_C \ll E_F$), for a constant T = 300 K and for the zero-current approximation. For comparison, a metal point source has been also simulated (F = 5 GV/m, R = 20 nm, ϕ = -4.5 eV, γ = 10, and T = 300 K).

The electron energy distribution of i-Ge (Fig. 4b) shows the characteristic band gap of the semiconductor, with the electrons from the conduction band in blue and from the valence band in orange. Both the conduction and valence distributions present a peak because of the combined effect of the density of states (DoS), tending to zero as it approaches $E_C$ and $E_V$, and the Fermi-Dirac distribution (Fig. 4a). The energy electron distribution of the metal emitter (green) peaks around the Fermi level, as a result of the combination of the Fermi-Dirac distribution and the transmission coefficient.

GETELEC-2.0 can be used to extract the potential barrier and the emitter surface temperature by fitting the right-hand side tail of $E_{V_{peak}}$ to the transmission coefficient D(E) and the left-hand side tail of $E_{C_{peak}}$ to the Fermi-Dirac distribution, respectively. The energy distribution between $E_{C_{peak}}$ and $E_{V_{peak}}$ yields information about the DoS, but such analysis goes beyond of remit of this manuscript and will be addressed in the future.

However, obtaining high-quality data for analysis is not trivial, especially for semiconductors. In fact, to the knowledge of the authors, there are no reports for the experimental observation of either $E_{C_{peak}}$ and $E_{V_{peak}}$ or the semiconductor band gap. Our model suggests (Fig. 5) reasons for such lack of experimental evidence, from which we can make suggestions for future experimental tests.

At high electric fields, the electrons accumulate at the emitter surface degenerating the semiconductor; point from which the emission from the semiconductor presents an energy distribution like that of a metal emitter (Fig. 5 green (metal) and blue (degenerate semiconductor) lines). This is consistent with the observation reported by Arthur on n-Ge [20]. However, it does not mean there is no emission from the valence band, but it has such a low magnitude in comparison with that of the emission from the valence band that it is unlikely an electron energy analyzer has the high dynamic range required to measure both contributions.

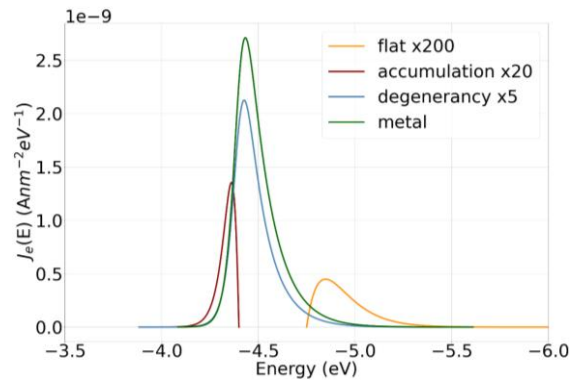

Figure 5: Energy distribution of a metal emitter (green), degenerate semiconductor (blue), electron accumulation (red) where one can expect semiconductor like distribution with conduction tail leading the emission, and flat band (orange) where the emission is led by the valence band. For visualisation purposes the electron counts have been multiplied by a factor ($\times factor$).

As the field is reduced, less charge is accumulated at the surface relaxing the bands and transitioning from degeneracy to electron accumulation. From this point the energy distribution is characteristic of that of a semiconductor with a sharp drop in the electron count (Fig. 5 red line) around the band gap. The electron emission is still led by the conduction band, with emission from the valence band being three orders of magnitude lower. The valence band component is the leading emission term on the flat band scenario. The orange line (Fig. 5) shows the characteristic shape of valence distribution with sharp drop to the left where the band gap is.

Figure 5 shows how the electron energy distribution goes from metal like, to conduction band led to valence band led as the field magnitude is reduced or the band bending relaxed. The existence of a leading component makes the measurement of both peaks and band gap of the electron energy distribution challenging because of the very wide dynamic range. Our model predicts a point at which the conduction and valence components have the same magnitude, which could be used to observe the band gap in the emission characteristics. Such point is discussed in subsection 4.2.

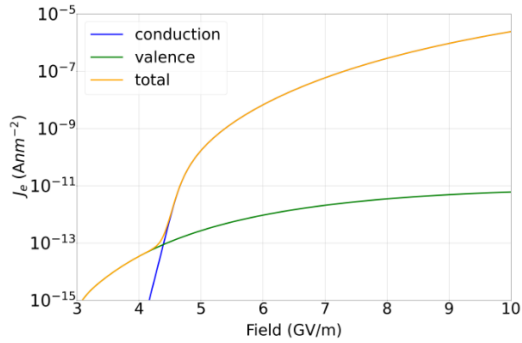

Figure 6: Dependence of band contribution to emission as a function of the applied field. Total emitted current density (orange), valence component (green) and conduction component (blue).

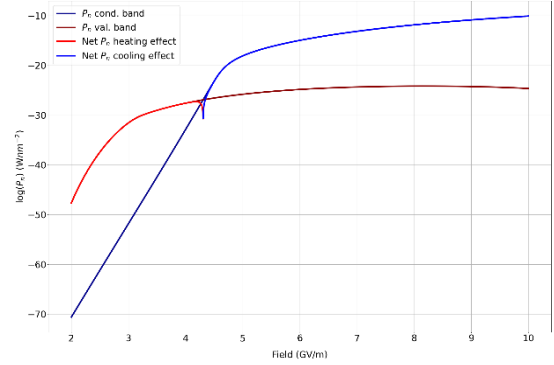

Figure 2: Evolution of the Nottingham heat as a function of the electric field. For low fields (E < 4 GV/m), where the valence band leads the emission, most electrons come from the below the $E_F$; thus, heating up the emitter. At higher fields most electrons come from above $E_F$, which has a net cooling effect.

## 2. Emission characteristics and saturation

As explained on the previous section, at low electric fields the bands are flat and the valence component to the emission leads the emission. With increasing fields, the valence component presents a slower growth compared to that of the conduction component. This slower growth is due to the fact that as the field increases and the potential barrier becomes thinner increasing the tunnelling probability, the valence band is pushed down, removing electrons from high energy states, therefore balancing both effects (Fig. 6 green line). On the other hand, as the conduction band is pushed down and approaches $E_F$ the number of electrons available for emission grows exponentially as per the Fermi-Dirac distribution and quickly taking over as main contributor to the total emitted current (Fig. 6 blue line).

Such exponential growth continues until the emitter surface degenerates due to the accumulation of charges at the semiconductor's surface. At this point, the emission will predominantly come from around the Fermi level, like a metal (Fig. 5), and any further increase on the field results on voltage drop on the semiconductor with the bands being pushed down and, thus, emission saturation.

The slope of the valence band component (green) corresponds to the transmission coefficient, which remains monotonic over the range of simulated values. As the band is pushed down into deeper energy levels, the electron energy decreases making G(E) to grow exponentially (see Eq. 3). Therefore, despite the fact there are electrons available at the valence band, their associated tunnelling probability quickly goes to zero and the emission component from the valence band vanishes. On the other hand, as the bands are pushed down the conduction component of the emission (blue) grows. This represents the fact that more electrons become available for energies at which the tunnelling probability is high. The slope of the conduction component is proportional to the Fermi Dirac distribution, which remains static at the Fermi level. At the point where $E_C < \sim 2E_F$, the emission predominately comes from $E_F$ and any further increase of the electric field will not yield more current. This saturation has been experimentally observed in p-type semiconductors [1]. Our model predicts that such saturation also occurs in n-type semiconductors, given that the emitter temperature remains constant, but no experimental evidence has been yet found.

The conduction component takes over the valence component at $E_C = E_F$, point from which the emission exponentially grows until its saturation. Before that point, a threshold should be surpassed until exponential growth is reached (Fig. 6). Such behaviour is analogous to that of a MOSFET [21].

## 3. Heat effects

The Nottingham heating effect is defined as the net deposited/extracted energy due to the energy difference between the emitted and replacing electrons. For metals with spherical isoenergetic Fermi surfaces, the average replacement electron energy is taken to be $E_F$, without any significant error [2]. For semiconductors, this is less clear since the replacement energy depends on the exact band structure and doping level at a given position on the junction at the other side of the cathode [22]. For our model, we set the replacement energy to be at $E_F$, under the assumption that the quasi-Fermi level in the semiconductors is always defined to be close to the position of the vast majority of the mobile electrons, which are the ones contributing to the current.

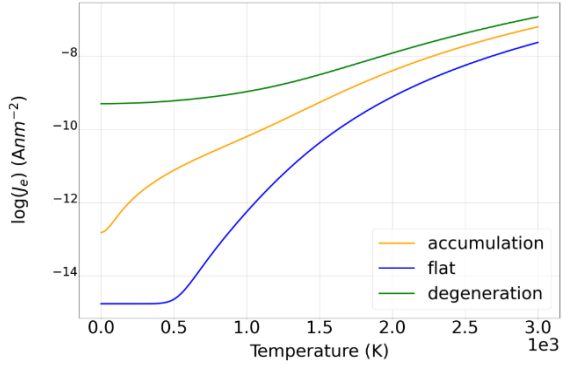

Figure 8: field emission curves as a function of the temperature for 3 band bending scenarios. The emission curves converge at high temperatures as the electrons obtain the energy to go "over" the barrier (thermionic emission) instead of "through" the barrier (field emission).

To show the dynamics of the Nottingham heat, we have plotted $P_N$ for different band scenarios as we did for the electron energy (Fig. 5). At low fields, the emission is led by the valence band. Low energy electrons are emitted, to be replaced by high energy electrons, thus depositing energy that increases the emitter's temperature. As the field increases and the bands bend downwards, the high energy electrons from the conduction band starts tunnelling out of the semiconductor and, with them, extracting energy from the system, which results in a net cool down of the emitter's surface. This picture is consistent with that reported by Fursey in metal emitters, where the emitter surface is several degrees cooler than the emitter's core [2].

## 4. Temperature dependence

Advances in field emission experimental setups have allowed to perform tests at cryogenic temperatures (< 50K) [23]. While metals at those temperatures behave like superconductors, semiconductors become insulators since no electrons have the energy enough to populate the conduction band. Our model enables the study of field emission from cryogenic to high temperatures.

We have plotted (Fig. 8) the emitted current for the flat-band, accumulation, and degeneracy for temperatures ranging from 0 to 3000 K. We also show (Fig. 9) the electron energy distribution for the same band same scenarios and varying temperatures (0, 300 and 1000 K).

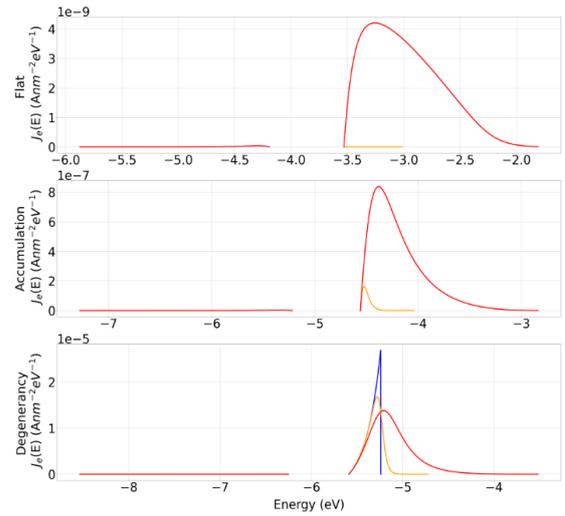

Figure 9: Electron energy distribution for flat, accumulation, and degenerate bands at 0, 300, and 1000 K.

Figure 8 shows how the characteristic emitted current converges, regardless of the band structure, as the emitter temperature rises. This is due to the fact that at higher temperatures more electrons become available at higher energies; thus, populating the states above $E_C$ and broadening the Fermi Dirac distribution. At that point, the electrons have enough energy to jump over the potential barrier, as expected in Schottky and thermionic emission. This is clearly seen in Fig. 9.

For the three band structures presented in Figure 9, we see that the main component of the emitted electron is that from the conduction band (red lines, which represent emission at 1000 K). For the flat and accumulation cases the left-hand side of the distribution tail quickly falls as $E_C$ is approached. While for the degeneracy case we see a smooth tail which is proportional to the potential barrier, since $E_C$ falls below $E_F$. The right-hand side is proportional to the Fermi-Dirac distribution at the calculated temperature. The differences in the order of magnitude for the three plots, represent the fact that at higher fields the potential barrier gets thinner, which enable electrons to tunnel through in addition of jumping over it. This picture is consistent with the previous literature and findings.

At moderate temperatures (~300 K) we observe the same electron distribution pattern that a high temperature. This is not the case from cryogenic temperatures (Fig. 9, blue line). At cryogenic temperatures, the electrons entirely come from the valence band, with a sharp drop at $E_V$ since there are not thermionically excited electrons above it, even though $E_C$ is well below $E_F$. The left-hand side of the tail still follows the same of the potential barrier.

## V. Conclusions

We have reviewed and corrected the mathematical expression for field emission from semiconductors. Our correction is for the electrons emitted from the valence band, which is of relevance at cryogenic temperatures.

GETELEC has been expanded to include the physics of field emission from semiconductors, both to calculate the emitted current density and Nottingham heat given inputs such as the band structure, the electric field, and temperature of the emitter. The calculations of the Nottingham heat provide a qualitative picture since our code cannot calculate the energy distribution of the replacement electrons.

To improve the model's computational efficiency, we have implemented a novel approach to evaluate the potential energy barrier by polynomial fitting. This approach has increased the speed of our code by factor of six, without sacrificing the numerical accuracy of the full JWKB method. We have also translated the code from Fortran to Python to make it more accessible and easier to use. A general tool for the calculation of field emission from semiconductors has been developed. The electron energy distribution of the emitted electrons can be calculated for a wide range of fields, temperatures, and band structures. Our model predicts the saturation of the emitted current density at high fields regardless of the doping levels.

Despite its versatility and usefulness, the model has limitations: (1) the model assumes a smooth emitter surface free of surface states, which might play an important role in the electron emission from semiconductors as reported by Modinos [12]. (2) The energy of the replacement electrons is set to be the Fermi energy as GETELEC cannot calculate it; for low field emission these results have to be interpreted qualitatively. (3) GETELEC provides the boundary condition for electron emission from semiconductors; to have a full picture of a semiconducting emitter our model would have to be coupled with multi-physics packages to solve the continuity and heat equations. These limitations will be the focus of future research.

*ACKOWLEDGEMENTS*

S.B.C acknowledges support from the Royal Society of Edinburgh (Saltire Fellowship Grant No. 1956).

*AUTHOR DECLARATIONS*

**No conflict of interest.** The authors have no conflicts of interest to disclose.

**REFERENCES**

[1] M. Choueib, R. Martel, C. S. Cojocaru, A. Ayari, P. Vincent, and S. T. Purcell, "Current saturation in field emission from H-passivated Si nanowires," *ACS Nano*, vol. 6, no. 8, pp. 7463–7471, Aug. 2012, doi: 10.1021/nn302744e.

[2] G. Fursey, *Field Emission in Vacuum Microelectronics*. 2003.

[3] S. B. Cárceles, A. Kyritsakis, V. Zadin, A. Mavalankar, and I. Underwood, "Field Emission Beyond Information Displays," in *Digest of Technical Papers - SID International Symposium*, John Wiley and Sons Inc, 2023, pp. 366–369. doi: 10.1002/sdtp.16568.

[4] M. E. Swanwick, P. D. Keathley, F. X. Kartner, and L. F. Velasquez-Garcia, "Ultrafast photo-triggered field emission cathodes using massive, uniform arrays of nano-sharp high-aspect-ratio silicon structures," *2013 Transducers and Eurosensors XXVII: The 17th International Conference on Solid-State Sensors, Actuators and Microsystems, TRANSDUCERS and EUROSENSORS 2013*, no. June, pp. 2680–2683, 2013, doi: 10.1109/Transducers.2013.6627358.

[5] G. Travish, F. J. Rangel, M. A. Evans, K. Schmiedehausen, and B. Hollister, "Addressable flat-panel x-ray sources for medical, security, and industrial applications," *Peocedings of Spie*, vol. 8502, no.

October 2012, 2012, doi: 10.1117/12.929354.

[6] J. V. Sci et al., "All field emission experiments are noisy, … are any meaningful? All field emission experiments are noisy, … are any meaningful?," vol. 024001, no. November 2022, 2023, doi: 10.1116/6.0002338.

[7] A. Kyritsakis and F. Djurabekova, "A general computational method for electron emission and thermal effects in field emitting nanotips," *Comput Mater Sci*, vol. 128, no. 4, pp. 15–21, Feb. 2017, doi: 10.1016/j.commatsci.2016.11.010.

[8] X. Gao et al., "evaporation and critical electric field of copper Molecular dynamics simulations of thermal evaporation and critical electric field of copper nanotips," 2020.

[9] R. Stratton, "Field emission from semiconductors," *Proceedings of the Physical Society. Section B*, vol. 68, no. 10, pp. 746–757, 1955, doi: 10.1088/0370-1301/68/10/307.

[10] R. Stratton, "Theory of Field Emission from Semiconductors," *Physical Review*, vol. 125, no. 1, pp. 67–82, 1962, doi: 10.1103/PhysRev.125.67.

[11] L. M. Baskin, O. I. Lvov, and G. N. Fursey, "General features of field emission from semiconductors," *Physica Status Solidi (B)*, vol. 47, no. 1, pp. 49–62, 1971, doi: 10.1002/pssb.2220470105.

[12] A. Modinos, *Field, Thermionic, and Secondary Electron Emission Spectroscopy*, vol. 102, no. 1. Boston, MA: Springer US, 1984. doi: 10.1007/978-1-4757-1448-7.

[13] A. Kyritsakis and J. P. Xanthakis, "Extension of the general thermal field equation for nanosized emitters," *J Appl Phys*, vol. 119, no. 4, 2016, doi: 10.1063/1.4940721.

[14] M. Choueib, A. Ayari, P. Vincent, M. Bechelany, D. Cornu, and S. T. Purcell, "Strong deviations from Fowler-Nordheim behavior for field emission from individual SiC nanowires due to restricted bulk carrier generation," *Phys Rev B Condens Matter Mater Phys*, vol. 79, no. 7, Feb. 2009, doi: 10.1103/PhysRevB.79.075421.

[15] R. Stratton, "Energy distributions of field emitted electrons," *Physical Review*, vol. 135, no. 3A, 1964, doi: 10.1103/PhysRev.135.A794.

[16] E. C. Kemble, "A Contribution to the Theory of the B. W. K. Method," *Physical Review*, vol. 48, Sep. 1935, doi: 10.1103/PhysRev.48.549.

[17] A. Kyritsakis and J. P. Xanthakis, "Derivation of a generalized Fowler-Nordheim equation for nanoscopic field-emitters," *Proceedings of the Royal Society A: Mathematical, Physical and Engineering Sciences*, vol. 471, no. 2174, 2015, doi: 10.1098/rspa.2014.0811.

[18] A. Kyritsakis, "General form of the tunneling barrier for nanometrically sharp electron emitters," *J Appl Phys*, vol. 133, no. 11, Mar. 2023, doi: 10.1063/5.0144608.

[19] K. L. Jensen, "General formulation of thermal, field, and photoinduced electron emission," *J Appl Phys*, vol. 102, no. 2, 2007, doi: 10.1063/1.2752122.

[20] J. R. Arthur, "ENERGY DISTRIBUTION OF FIELD EMISSION FROM GERMANIUM," North-Holland Publishing Co, 1964.

[21] S. M. Sze and K. K. Ng, *Physics of semiconductor devices*. Wiley-Interscience, 2007.

[22] M. S. Chung, Y. J. Jang, A. Mayer, P. H. Cutler, N. M. Miskovsky, and B. L. Weis, "Energy exchange in field

emission from semiconductors," *Journal of Vacuum Science & Technology B: Microelectronics and Nanometer Structures Processing, Measurement, and Phenomena*, vol. 26, no. 2, pp. 800–805, Mar. 2008, doi: 10.1116/1.2822944.

[23] M. Jacewicz, J. Eriksson, R. Ruber, S. Calatroni, I. Profatilova, and W. Wuensch, "Temperature-Dependent Field Emission and Breakdown Measurements Using a Pulsed High-Voltage Cryosystem," *Phys Rev Appl*, vol. 14, no. 6, Dec. 2020, doi: 10.1103/PhysRevApplied.14.061002.

## APPENDIX – Mathematical development of field emission equations.

1. Emission equations for conduction band electrons

$$j_C = \frac{q}{4\pi^3 \hbar^3} \int f_{FD}(E) D(E - E_r) \frac{\partial E}{\partial p_z} d^3p \quad (A.1)$$

where

$$E = \frac{p^2}{2m^*} \quad (A.2)$$

$$E_r = \frac{p_x^2 + p_y^2}{2m} = \frac{p_r^2}{2m} \quad (A.3)$$

where

m is the free electron mass of the electron

m$^*$ is the effective mass of the electron

We first change to polar coordinates

$$p_r^2 = p_x^2 + p_y^2 \quad (A.4)$$

$$\varphi = \tan^{-1} \frac{p_x}{p_y} \quad (A.5)$$

$$p_x = p_r \cos \varphi \quad (A.6)$$

$$p_y = p_r \sin \varphi \quad (A.7)$$

$$p_z = p_z \quad (A.8)$$

$$d^3p = p_r dp_r dp_z d\varphi \quad (A.9)$$

With all the above definitions

$$j_C = \frac{q}{4\pi^3 \hbar^3 m^*} \int_{-\infty}^{\infty} \int_0^{\infty} \int_0^{2\pi} f_{FD}(E) D(E - E_r) p_z p_r dp_r dp_z d\varphi \quad (A.10)$$

Since neither E nor $E_r$ depend on $\pi$

$$j_C = \frac{q}{2\pi^2 \hbar^3 m^*} \int_{-\infty}^{\infty} \int_0^{\infty} f_{FD}(E) D(E - E_r) p_z p_r dp_r dp_z \quad (A.11)$$

Making a change of variables

$$\{p_z, p_r\} \rightarrow \{E, E_r\} \quad (A.12)$$

$$|Jac| = \begin{vmatrix} \frac{\partial E}{\partial p_z} & \frac{\partial E}{\partial p_r} \\ \frac{\partial E_r}{\partial p_z} & \frac{\partial E_r}{\partial p_r} \end{vmatrix} = \begin{vmatrix} \frac{p_z}{m^*} & \frac{p_r}{m^*} \\ 0 & \frac{p_r}{m} \end{vmatrix} = \frac{p_z p_r}{m^* m} \quad (A.13)$$

$$p_z p_r dp_r dp_z = m^* m dE dE_r \quad (A.14)$$

Calculating the new integration limits

$$E = \frac{p_r^2 + p_z^2}{2m^*} = \frac{m E_r}{m^*} + \frac{p_z^2}{2m^*} \quad (A.15)$$

Since

$$\frac{p_z^2}{2m^*} \geq 0 \quad \rightarrow \quad E \geq \frac{m}{m^*} E_r \quad \rightarrow \quad E_r \leq \frac{m^*}{m} E = \alpha E \quad (A.16)$$

where

$$\alpha = \frac{m^*}{m} \quad (A.17)$$

So now we have

$$j_C = \frac{mq}{2\pi^2 \hbar^3} \int_0^{\infty} \int_0^{\alpha E} f_{FD}(E) D(E - E_r) dE_r dE \quad (A.18)$$

Now we define

$$E_z = E - E_r \quad \rightarrow \quad dE_z = -dE_r \quad (A.19)$$

$$\text{for:} \quad E_r = 0 \quad \rightarrow E_z = E \quad (A.20)$$

$$\text{for:} \quad E_r = \alpha E \quad \rightarrow E_z = \bar{\alpha} E \quad (A.21)$$

where

$$\bar{\alpha} = (1 - \alpha) \quad (A.22)$$

$$L = \frac{mq}{2\pi^2 \hbar^3} \quad (A.23)$$

So

$$j_C = L \int_0^\infty f_{FD}(E) \int_E^{\overline{\alpha}E} -D(E_z) dE_z dE$$

$$= L \int_0^\infty f_{FD}(E) \int_{\overline{\alpha}E}^E D(E_z) dE_z dE \quad (A.24)$$

$$j_C = L \int_0^\infty -l'_{FD}(E) g(E) dE \quad (A.25)$$

where

$$g(E) = \int_{\overline{\alpha}E}^E D(E_z) dE_z \quad (A.26)$$

Solving by parts

$$j_C = L \left[ [l_{FD}(E) g(E)]_0^\infty + \left[ \int_0^E l_{FD}(E) g'(E) dE \right] \right] (A.27)$$

$$g'(E) = D(E) - \overline{\alpha} D(\overline{\alpha}E) \quad (A.28)$$

and since

$$[l_{FD}(E) g(E)]_0^\infty = 0 \quad (A.29)$$

We get

$$j_C = L \int_0^E l_{FD}(E) \big(D(E) - \overline{\alpha} D(\overline{\alpha}E)\big) dE \quad (A.30)$$

From the later development, we can obtain the energy distribution of the electrons coming from the conduction band.

$$J_C = f_{FD}(E) \int_{\overline{\alpha}E}^E D(E_z) dE_z \quad (A.31)$$

2. Emission Equations for valence band equations

$$j_V = \frac{q}{4\pi^3 \hbar^3} \int f_{FD}(\overline{E}) D_V(\overline{E} + \overline{E}_r) \frac{\partial \overline{E}}{\partial p_z} d^3 p \quad (A.32)$$

where

$$\overline{E} = E_v - \varepsilon \quad (A.33)$$

$$\varepsilon = E_V - E = E_V - \frac{p^2}{2m^*} = E_V - \frac{p_r^2 + p_z^2}{2m^*} \quad (A.34)$$

$$p_r^2 = p_x^2 + p_y^2 \quad (A.35)$$

$$\overline{E}_r = \frac{p_r^2}{2m} \quad (A.36)$$

and with

$$D_V(E) = D(E_V - E) \quad (A.37)$$

we get

$$D_V(\overline{E} + \overline{E}_r) = D_V\left(E_V - E + \frac{p_r^2}{2m}\right)$$

$$= D\left(E_V - E_V + E - \frac{p_r^2}{2m}\right)$$

$$= D\left(E - \frac{p_r^2}{2m}\right) \quad (A.38)$$

Following the same reason for $f_{FD}(\overline{E})$ we get

$$f_{FD}(\overline{E}) = f_{FD}(E) \quad (A.39)$$

where

m is the free electron mass of the electron

m* is the effective mass of the hole

We all the above definitions we obtain

$$j_V = \frac{q}{4\pi^3 \hbar^3 m^*} \int f_{FD}(E) D\left(E - \frac{p_r^2}{2m}\right) p_z d^3 p \quad (A.40)$$

Changing to polar coordinates, as done for $j_c$, we get

$$j_V = \frac{q}{2\pi^2 \hbar^3 m^*} \int_{-\infty}^\infty \int_0^\infty f_{FD}(E) D\left(E - \frac{p_r^2}{2m}\right) p_z p_r dp_r dp_z \quad (A.41)$$

Making a change of variables

$$\{p_z, p_r\} \to \{E, E_r\} \quad (A.42)$$

$$|Jac| = \begin{vmatrix} \frac{\partial E}{\partial p_z} & \frac{\partial E}{\partial p_r} \\ \frac{\partial E_r}{\partial p_z} & \frac{\partial E_r}{\partial p_r} \end{vmatrix} = \begin{vmatrix} -\frac{p_z}{m^*} & -\frac{p_r}{m^*} \\ 0 & \frac{p_r}{m} \end{vmatrix} = -\frac{p_z p_r}{m^* m} (A.43)$$

$$p_z p_r dp_r dp_z = -m^* m dE dE_r \quad (A.44)$$

Calculating the new integration limits

$$E = E_V - \frac{m E_r}{m^*} + \frac{p_z^2}{2m^*} \quad (A.45)$$

Since

$$E_r \geq 0 \quad \to \quad E \leq E_V - \frac{m}{m^*} E_r$$

$$\to \quad E_r \leq \frac{m^*}{m}(E_V - E) = \alpha(E_V - E) \quad (A.46)$$

where

$$\alpha = \frac{m^*}{m} \quad (A.47)$$

So now we have

$$j_V = \frac{mq}{2\pi^2 \hbar^3} \int_{-\infty}^{E_V} \int_0^{\alpha(E_V-E)} f_{FD}(E)D(E-E_r)dE_r dE \quad (A.48)$$

Now we define

$$E_z = E - E_r \quad \rightarrow \quad dE_z = -dE_r \quad (A.49)$$

$$\text{for: } E_r = 0 \quad \rightarrow E_z = E \quad (A.50)$$

$$\text{for: } E_r = \alpha(E_V - E) \quad \rightarrow E_z = \bar{\alpha}E - \alpha E_V \quad (A.51)$$

where

$$\bar{\alpha} = (1 + \alpha) \quad (A.52)$$

$$L = \frac{mq}{2\pi^2 \hbar^3} \quad (A.53)$$

So

$$j_V = L \int_{-\infty}^{E_V} f_{FD}(E) \int_{\bar{\alpha}E-\alpha E_V}^{E} D(E_z) dE_z dE \quad (A.54)$$

$$j_V = L \int_{-\infty}^{E_V} -l'_{FD}(E) g(E) dE \quad (A.55)$$

where

$$g(E) = \int_{\bar{\alpha}E-\alpha E_V}^{E} D(E_z) dE_z \quad (A.56)$$

Solving by parts

$$j_V = L \left[ [-l_{FD}(E)g(E)]_{-\infty}^{E_V} + \left[ \int_{-\infty}^{E} l_{FD}(E) g'(E) dE \right] \right] (A.57)$$

$$g'(E) = D(E) - \bar{\alpha} D(\bar{\alpha}E - \alpha E_V) \quad (A.58)$$

and since

$$[-l_{FD}(E)g(E)]_{-\infty}^{E_V} = 0 \quad (A.59)$$

We get (with $E_V$ as upper limit)

$$j_V = L \int_{-\infty}^{E_V} l_{FD}(E)(D(E) - \bar{\alpha} D(\bar{\alpha}E - \alpha E_V)) dE (A.60)$$

which contrasts with Stratton's

$$j_V = L \int_{-\infty}^{E_V} l_{FD}(E)(D(E) - \bar{\alpha} D(\bar{\alpha}E - \alpha E_V)) dE + \ln\left\{1 + e^{\frac{E_V}{kT}}\right\} \int_{E_V}^{\bar{\alpha}} D(E) dE (A.61)$$

3. From the later development, we can obtain the energy distribution of the electrons coming from the conduction band.

$$J_V = f_{FD}(E) \int_{\bar{\alpha}E-\alpha E_V}^{E} D(E_z) dE_z \quad (A.62)$$